\documentclass[12pt]{article}
\usepackage{times}
\usepackage{amssymb}
\usepackage{graphicx}
\usepackage{color}

\newcommand{\nc}{\newcommand}
\newcommand{\rnc}{\renewcommand}

\rnc{\baselinestretch}{1.24}    % 1.5 spacing btwn text lines
\rnc{\arraystretch}{1.24}       % spacing btwn the rows of a non-eqn array

\makeatletter
\rnc{\theequation}{\thesection.\arabic{equation}}
\@addtoreset{equation}{section}
\makeatother                      

\nc{\be}{\begin{eqnarray}}
\nc{\ee}{\end{eqnarray}}
\nc{\xx}{\nonumber\\}

\nc{\eq}[1]{(\ref{#1})}
\nc{\newcaption}[1]{\centerline{\parbox{6in}{\caption{#1}}}}

\nc{\fig}[3]{
\begin{figure}
\centerline{\epsfxsize=#1\epsfbox{#2.eps}}
\newcaption{#3. \label{#2}}
\end{figure}
}

\nc{\np}[3]{Nucl. Phys. {\bf B#1} (#2) #3}
\nc{\pl}[3]{Phys. Lett. {\bf #1B} (#2) #3}
\nc{\prl}[3]{Phys. Rev. Lett.{\bf #1} (#2) #3}
\nc{\prd}[3]{Phys. Rev. {\bf D#1} (#2) #3}
\nc{\ap}[3]{Ann. Phys. {\bf #1} (#2) #3}
\nc{\prep}[3]{Phys. Rep. {\bf #1} (#2) #3}
\nc{\rmp}[3]{Rev. Mod. Phys. {\bf #1} (#2) #3}
\nc{\cmp}[3]{Comm. Math. Phys. {\bf #1} (#2) #3}
\nc{\mpl}[3]{Mod. Phys. Lett. {\bf #1} (#2) #3}
\nc{\cqg}[3]{Class. Quant. Grav. {\bf #1} (#2) #3}
\nc{\jhep}[3]{J. High Energy Phys. {\bf #1} (#2) #3}

\def\CO{{\cal O}}

\def\IR{\mathbb{R}}
\def\IZ{\mathbb{Z}}

\def\a{\alpha}
\def\b{\beta}

\def\d{\delta}

\def\l{\lambda}
\def\m{\mu}
\def\n{\nu}
\def\th{\theta}

\def\s{\sigma}

\def\D{\Delta}
\def\G{\Gamma}

\def\half{\frac{1}{2}}

\def\goto{\rightarrow}

\def\del{\nabla}

\newcommand{\bra}[1]{\langle{#1}|}
\newcommand{\ket}[1]{|{#1}\rangle}

\def\Tr{{\rm Tr}}

\def\ao{{\alpha_1}}
\def\bo{{\beta_1}}
\def\at{{\alpha_2}}
\def\bt{{\beta_2}}
\def\aod{{\dot{\alpha_1}}}
\def\bod{{\dot{\beta_1}}}
\def\atd{{\dot{\alpha_2}}}
\def\btd{{\dot{\beta_2}}}

\begin{document}
%%%%%%%%%%%%%%%%%%%%%%%%%%%%%%%%%%%%%%%%%%%%%%%%%%%%%%%%%%%%%%%%%
%                                                               %
%                       TITLE PAGE      
%                                                               %
%%%%%%%%%%%%%%%%%%%%%%%%%%%%%%%%%%%%%%%%%%%%%%%%%%%%%%%%%%%%%%%%%

\begin{titlepage}

%\hfill{QMUL-PH-04-07}

\begin{flushright}
hep-th/0409261 \\
CERN-PH-TH/2004-188 \\
QMUL-PH-04-07
\end{flushright}
\vspace*{2.0cm}
\centerline{\Large\bf Holographic cubic vertex in the pp-wave}
%\vspace*{0.5cm}
%\centerline{\Large\bf }
\vspace*{1.5cm} 
\centerline{Sangmin Lee and Rodolfo Russo\footnote{On leave of absence from {\em Queen Mary, University of London},  E1 4NS London, UK.}
}
\vspace*{1.0cm}
\centerline{\sl }
\centerline{\sl CERN, CH-1211 Geneva 23, Switzerland}
\centerline{\sl }
\vskip0.3cm
\vspace*{2.0cm}
\centerline{\bf ABSTRACT}
\vspace*{0.5cm}
\noindent
We revisit the cubic interaction of IIB string theory in the maximally
supersymmetric pp-wave background. In the supergravity limit, we show
that detailed comparison with AdS supergravity determine the vertex
completely. Extension of this supergravity vertex to the full string
theory leads to a new cubic vertex that combines the previous
proposals and contains additional terms. We give an alternative
derivation of the holographic duality map in supergravity, first found
by Dobashi and Yoneya (hep-th/0406225) and show that our new vertex is
consistent with it. We compare some non-BPS amplitudes (including
impurity non-preserving ones) with the corresponding field theory
correlators, and discuss what they imply for the stringy
generalization of the duality map.  We also notice that our vertex
realizes the $U(1)_Y$ symmetry linearly, and propose a similar
modification for the flat space vertex.

\vspace*{1.1cm}

\end{titlepage}
\setcounter{footnote}{0}

%%%%%%%%%%%%%%%%%%%%%%%%%%%%%%%%%%%%%%%%%%%%%%%%%%%%%%%%%%%%%%%%%
%                                                               %
%       SECTION 1. INTRODUCTION 
%                                                               %
%%%%%%%%%%%%%%%%%%%%%%%%%%%%%%%%%%%%%%%%%%%%%%%%%%%%%%%%%%%%%%%%%

\section{Introduction} 

The BMN duality~\cite{Berenstein:2002jq} has drawn a lot of attention
for the past two years, largely because it opened up a systematic way
to test the AdS/CFT correspondence~\cite{Maldacena:1997re} at the
string level. 
The most striking discovery was that the tree-level string
spectrum~\cite{Metsaev:2001bj,Metsaev:2002re} in the maximally
supersymmetric pp-wave background~\cite{Blau:2001ne,Blau:2002dy}
matches exactly (that is, to all orders in the $\a'$-expansion), a
particular class of ${\cal N}=4$ super Yang-Mills
operators~\cite{Berenstein:2002jq}. Since then, much effort has been
made to understand how the string interactions (non-zero $g_s$) fit
into the duality. In spite of many important works in the
literature\footnote{See the review papers
\cite{Pankiewicz:2003pg,Plefka:2003nb,Maldacena:2003nj,Spradlin:2003xc,Sadri:2003pr,Russo:2004kr}
for a detailed bibliography.}, the problem has not been fully solved
yet.  The goal of this paper is to report some progress on this
subject.

The simplest type of string interaction is the cubic interaction, 
in which two strings join to form a single string or vice versa.
There are three crucial issues concerning the cubic interaction 
in the pp-wave duality.
\begin{enumerate} 

\item 
Construction of the cubic vertex.

The string theory in the pp-wave is formulated in terms of the
Green-Schwarz superstring in the light-cone gauge.  In this set-up,
the $3$-string vertex is given by the cubic part of the light-cone
Hamiltonian.  The vertex is usually constructed by imposing the
super-symmetry constraints. However, unlike in flat-space, the
constraints do not completely fix the pp-wave vertex.

\item 
Holographic duality map. 

Once the cubic Hamiltonian is known, one can compute its matrix
elements and obtain the coupling among three arbitrary string states.
On the Yang-Mills side, the natural observable is the coefficient of
the (normalized) cubic correlator. To make the comparison between
these two observables, one needs a duality map, which must somehow
`know' about the holography underlying the original AdS/CFT
correspondence.

\item 
Choice of basis (Operator mixing)

It is important to understand how the string and the
Yang-Mills Hilbert spaces are mapped to each other.
While the matching of the free spectra focuses mainly on the
eigenvalues of the physical observables, the duality map for the cubic
interaction tests in a much stronger way the dictionary between string
and gauge theory states. 
\end{enumerate} 
In this paper, we will discuss some new findings and considerations 
on these three points.

Spradlin and Volovich~\cite{Spradlin:2002ar,Spradlin:2002rv} made the
first proposal for the cubic vertex, which was further elaborated
in~\cite{Pankiewicz:2002gs,Pankiewicz:2002tg,Pankiewicz:2003kj,Pankiewicz:2003ap}.
Aside from satisfying the pp-wave super-algebra, the SV vertex has two
features: (a) it has definite parity under the accidental $\IZ_2$
symmetry that exchanges the two manifest $SO(4)$ symmetry groups (the
parity is odd in the conventions where the vacuum is $\IZ_2$
invariant), (b) it has a smooth `flat space' limit.  Before the
question of whether these features are compatible with the putative
duality map was answered, another physically different vertex was
proposed in~\cite{Chu:2002eu,Chu:2002wj,DiVecchia:2003yp}. This vertex
satisfies the same pp-wave super-algebra, but does not share the
above-mentioned features: (a) it has opposite parity under the
$\IZ_2$, (b) as a consequence of this parity property, it does not have
a smooth `flat space' limit.

Which one of the two vertices is the correct one? 
In fact, since the constraint from super-algebra essentially gives 
a set of linear differential equations, the right question would be  
``Which linear combination of the two is the correct one?'' 
Moreover, there may even exist other independent solutions to the 
super-algebra equations, ending up with a multi-dimensional space of 
candidate vertices. 

Clearly, to resolve the situation, one has to understand better how
holography works in the pp-wave. Among others, Yoneya and
collaborators have pursued this line of thought
systematically~\cite{Dobashi:2002ar,Yoneya:2003mu}. Recently,
in~\cite{Dobashi:2004nm}, they derived an explicit holographic duality
map for the supergravity sector of the pp-wave string theory by taking
the semi-classical limit of the GKPW
relation~\cite{Gubser:1998bc,Witten:1998qj} in AdS/CFT. This map led
them to conclude that the correct vertex is a particular linear
combination of the two vertices introduced above which breaks the
$\IZ_2$ symmetry `maximally'.

In this paper, we first re-derive the same duality map from a somewhat
different perspective, following the idea which first appeared in
\cite{Kiem:2002xn}. 
Then, we take a closer look at what it implies for the cubic
vertex. Among other things, we pay attention to the $U(1)_Y$ symmetry
of supergravity as well as the matrix elements of the
super-descendants of the chiral primary state. We find that the
proposal of~\cite{Dobashi:2004nm} should be further modified to
include three new terms similar to the second vertex mentioned above,
in order for the duality map to hold.  Our derivation indicates that
this vertex is the unique one compatible with the duality map, although
a rigorous proof is not yet available.  Finally, we discuss how to
extend the duality map to the full string theory. Suggestive as our
computation of stringy amplitudes are, the final answer seems to
require more work including sub-leading order computations in
Yang-Mills.

This paper is organized as follows. Sections~2 and~3 focus on
supergravity (or BPS) processes. Section~2 contains the derivation of
the holographic duality map. In section~3, we first derive a number of
AdS$_5 \times S^5$ $3$-point couplings and study their large $J$
limit. Then we discuss the $U(1)_Y$ symmetry of type IIB supergravity
and use it as an additional constraint on the pp-wave cubic vertex. A
unique answer for this vertex is obtained by requiring that it
reproduce the large $J$ limit of the previously derived AdS$_5 \times
S^5$ $3$-point couplings.
In section~4, we go beyond the supergravity sector and study the cubic
interaction among generic string states. In our construction, we
demand that the zero-mode structure of the string vertex reproduce
the supergravity results derived in the previous section. By combining
the known vertices and also adding some new terms, we present a
consistent proposal for the holographic $3$-string vertex. In order to
test its validity, we compute some stringy amplitudes and compare them
against the field theory results by using the simplest generalization
to the full string theory of the duality map. Section~5 contains our
conclusions along with a discussion of possible future directions.

\section{Holography in supergravity}

The holographic duality map in the supergravity sector 
can be derived in two simple steps\footnote{The main ideas of this section 
were first discussed in a limited setting 
in the appendix of Ref.~\cite{Kiem:2002xn}.}.
The first step is to note that the interaction part of the pp-wave 
Hamiltonian is equal to that of the original AdS geometry 
in the Penrose limit. 
This relation is not restricted to the BPS sector, 
but should hold even for the full string theory. 
The second step is to relate the AdS Hamiltonian 
to the coefficients of the gauge theory correlators 
via the GKPW relation in supergravity~\cite{Gubser:1998bc,Witten:1998qj}. 
This is possible since both quantities can be obtained 
from the same IIB supergravity action on AdS$_5\times S^5$. 

\subsection{From AdS to pp-wave}

The first step is a direct consequence of the standard AdS/CFT and 
pp-wave dictionaries. In the following table, we summarize in the
first two columns the two parameters that define each theory 
and define the dimensionless Hamiltonians in the third column.

\vskip 5mm

\begin{tabular}{rccc}
    & YM-loop / stringy effect & genus / string loop & Hamiltonian \\
AdS & $\l = g_{YM}^2 N = (R_{AdS}/l_s)^4$ & $1/N$ &  
$H^{\rm (AdS)} \equiv R_{AdS} P_0 $ 
\\
PP  & $\l' = g_{YM}^2 N / J^2 = 1/(\mu p^+\a')^2$ & $g_2= J^2/N$ & 
$H^{\rm (PP)} \equiv P_+/\mu $ 
\end{tabular}

\vskip 5mm

\noindent
Since the pp-wave theory describes the dynamics of AdS$_5 \times S^5$
in the Penrose limit ($N, J\to\infty$, keeping $\lambda'$ and
$g_2$ fixed), then the two Hamiltonians must be the same in this limit, 
except for the shift by $J$, which changes only the {\em free} part: 
\be
\label{inth}
\fbox{
$\lim\limits_{{\rm Penrose}} 
\left\{ H^{\rm (AdS)}(\lambda,1/N) -J \right\} =
H^{\rm (PP)}(\lambda',g_2) 
%(J\goto \infty) 
$
} ~.
\ee
In passing, we should emphasize that the mass scale $\mu$ in the
pp-wave has absolutely no physical meaning. The expressions such as
`$\mu \goto 0$' or `$\mu \goto \infty$' often found in the literature
should be interpreted as large $\l'$ and small $\l'$, respectively.
In particular, the bona-fide flat space ($\mu =0$) is not related to
the `$\mu \goto 0$' limit of the pp-wave.  If the two were smoothly
connected, the BMN duality would imply a holographic relation between
IIB string theory in flat space and a very strongly coupled gauge
theory.  Of course, some ingredients (for example, the Neumann matrix)
of the pp-wave Hamiltonian formally have a smooth $\mu \goto 0$ limit.
However, as we will see in the next section, the cubic Hamiltonian
contains pieces which manifestly break symmetries of flat space.
Discontinuity of the `$\mu \goto 0$' limit has been recently noticed
also in~\cite{Chu:2004iv}, where the causality properties of the
pp-wave string theory are studied.

Note also that we are taking the Penrose limit on the AdS Hamiltonian. 
This is to be contrasted with the approach of \cite{Kiem:2002pb}, where 
the Hamiltonian is computed directly in the pp-wave geometry.

\subsection{Hamiltonian vs. Correlator} 

Now we move on to the second step of the derivation. 
Suppose we have primary operators $O_i(x)$ in a CFT$_d$ and 
the corresponding scalar fields $\varphi_i$ living in AdS$_{(d+1)}$. 
Assume that the bulk action takes the standard form,
\be
S = - \int d^{d+1}x \sqrt{-g} \left[ \half (\del \varphi^i)^2 + 
\half m_i^2 (\varphi^i)^2 + \frac{1}{6} G_{ijk}^c 
\varphi^i \varphi^j \varphi^k \right] , 
\ee
where the AdS mass of $\varphi_i$ and the scaling dimension of $O_i$
are related by $m^2 = \D (\D - d/2)$. The superscript in $G_{ijk}^c$
is to stress that in this section we are working with canonically
normalized fields.

There are two things we can do with this action. First, we 
can compute in supergravity the normalized 3-point 
correlators following~\cite{Witten:1998qj,Freedman:1998tz},  
\be
&&\langle O_1(x_1) O_2(x_2) O_3(x_3)\rangle = 
\frac{C_{123}}{|x_1-x_2|^{2\b_3}|x_2-x_3|^{2\b_1} |x_3-x_1|^{2\b_2}}~, \\
\label{ccc}
&&C_{123} = \frac{G_{123}^c}{2^{5/2} \pi^{d/4}} \times
\prod_{r=1}^3 
\left( \frac{\G(\b_r)}{\{\G(\D_r-d/2+1)\G(\D_r)\}^{1/2}} \right) \times 
\G(\s - d/2)~,
\ee
where $\s = (\D_1 + \D_2 +\D_3)/2$, $\b_r = \s -\D_r$. 
Second, we can canonically quantize the free part of the action and 
read off the matrix elements of the cubic Hamiltonian. 
As usual, canonical quantization associates a harmonic oscillator to each 
normalizable solution to the free field equation of motion. 
For a real scalar in AdS, the expansion takes the following form:
\be
\label{cano}
\varphi(t, x) = \sum_i \frac{f_i(x)}{\sqrt{2(\D +n_i)}} 
\left( a_i e^{-i(\Delta + n_i)t} 
+ a^\dagger_i e^{i(\Delta + n_i)t} \right)~, 
\ee
where $t$ is the global time and $x$ denotes the $d$ spatial coordinates 
in the metric, 
\be
\label{glomet}
ds^2 = \frac{1}{\cos^2\th}
\left( - dt^2 + d\th^2 + \sin^2\th d\Omega_{d-1}^2 \right) .
\ee
In \eq{cano}, the index $i$ in runs over all solutions, 
and $f_i(x)$ are the spatial part of the solutions. 
The excitation number $n_i$ is zero 
for the ground state and is a positive integer for excited states. 
The matrix elements of the cubic Hamiltonian can be read off 
simply by inserting \eq{cano} into the cubic term of the Hamiltonian. 
For the ground state wave functions of the scalars, 
\be
f_0 = \sqrt{\frac{\G(\D+1)}{\pi^{d/2}\G(\D-d/2+1)}}(\cos\th)^\D~,
\ee
the matrix elements turn out to be\footnote{
Excited states give different values 
of $H_{123}$ through the overlap integral of wave-functions. 
However, note that all the wave-functions of a same field 
share the coupling constant $G^c_{123}$. 
This fact will be important in section 3. 
},
\be
\label{adsham}
H_{123} = \frac{G_{123}^c}{2^{3/2} \pi^{d/4}} \times
\prod_{r=1}^3 \left( \frac{\G(\D_r)}{\G(\D_r-d/2+1)} \right)^{1/2}
\times \frac{\G(\s - d/2)}{\G(\s)}~.
\ee
Comparing (\ref{ccc}) and (\ref{adsham}), one finds that 
\be
H_{123} = 
\frac{ 2\G(\D_1)\G(\D_2)\G(\D_3) }{ \G(\b_1)\G(\b_2)\G(\b_3)\G(\s) } 
C_{123} , 
\ee
In the pp-wave limit, we take $\Delta_r \goto \infty$ and use the 
relation (\ref{inth}) to obtain the holographic duality map as advertised,
\be
\label{hmap}
\fbox{$ 
\displaystyle
H^{({\rm PP})}_{123} = 
\lim\limits_{{\rm Penrose}} H^{({\rm AdS})}_{123}  = 
\frac{\D_{123}}{(\D_{123}/2)!} 
\left(\frac{J_1 J_2}{J_3} \right)^{\frac{\D_{123}}{2}} C_{123}
$}~,
\ee
where $\Delta_{123} \equiv \D_1 + \D_2 - \D_3$ is kept finite. 
From here on, for any physical quantity $X$ assigned to each of the 
three states participating in the cubic interaction, 
we will use the notation $X_{123} \equiv X_1 + X_2 - X_3$. 

\subsection{Intuitive picture by Yoneya et al.} 

Holography in the pp-wave duality has remained a puzzle 
because the boundary of the original AdS is completely lost 
in the process of taking the pp-wave limit.  
Then, how can one derive a relation like (\ref{hmap}) from 
the pp-wave string theory (or supergravity) without 
tracing back to the original AdS? Perhaps one cannot. 
We did trace back to the original AdS to derive (\ref{hmap}). 
In the next section, we will use it as a dynamical {\em input} 
in constructing the cubic vertex in the pp-wave. 
In other words, among all candidate vertices 
satisfying the (super-)symmetry constraints, 
we will pick out the 
one respecting the duality map.
This point of view was pursued systematically by Yoneya and
collaborators~\cite{Dobashi:2002ar,Yoneya:2003mu,Dobashi:2004nm}. We
briefly review their work here from a slightly different perspective.
It will provide an intuitive understanding of what (\ref{hmap}) means.

\begin{figure}
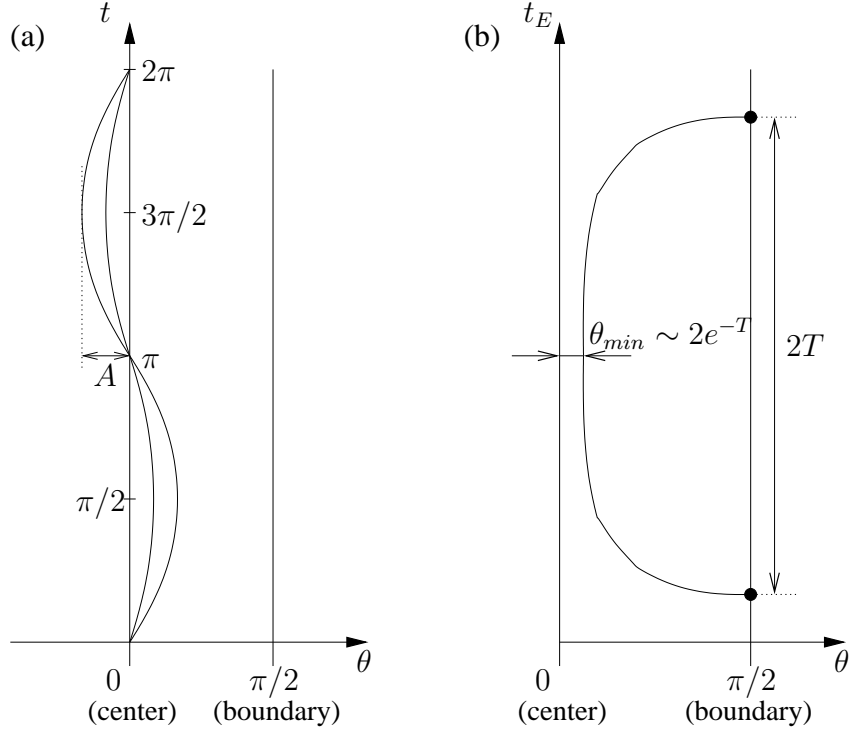
 
\begin{center}
\input geodesic.pstex_t
\end{center}
\caption{The geodesics in (a) Lorentzian and (b) Euclidean AdS in global  
coordinates.}
\end{figure}

One starts with the GKPW relation for the correlators. 
As emphasized in \cite{Dobashi:2002ar,Yoneya:2003mu,Dobashi:2004nm}, 
the bulk to boundary propagator should be understood 
as a Euclidean path integral. 
The reason is that in the Lorentzian signature, a massive 
particle can never reach the boundary. 
In global coordinates with the metric \eq{glomet},
the geodesic equation can be easily solved. 
For example, the solutions describing a radial motion look like 
(See Figure~1), 
\be
{\rm Lorentzian}: \;\;  \sin\th = \sin A \sin t ~, \;\;\;\;\;\;\;\;\;
{\rm Euclidean}: \;\;\sin\th = \frac{\cosh t}{\cosh T}~.
\ee
The second thing to notice is that in the semiclassical limit ($\D \gg
1$), the saddle point approximation to the 'propagator' along the
geodesic becomes reliable.  For a large value of the distance $2T$ in
time direction between the two boundary points, the Euclidean geodesic
starting from a boundary point runs exponentially toward the center of
the AdS and stays there until it curves back to the other boundary
point. 
This is consistent with the fact that the pp-wave limit
magnifies the small region around the center. 
In~\cite{Dobashi:2004nm}, the duality map (\ref{hmap}) was
derived by systematically performing the saddle point approximation
and constructing an effective action for a particle along the geodesic.  

Writing (\ref{hmap}) as $C_{123} = L_{123} \cdot H_{123}$, 
one could heuristically argue that $H_{123}$ originates from 
the body of the geodesic passing through
the center of AdS captured by the pp-wave, 
while $L_{123}$ comes from the `legs' connecting the center and the boundary. 
It would be very interesting to generalize this semi-classical picture 
to the full string theory and derive a similar duality map. 

\section{Supergravity vertex}

In this section, we derive the supergravity vertex 
consistent with the duality map (\ref{hmap}), 
leaving the full string theory vertex to the next one. 
In the first subsection, we compute several examples of 
$H_{123}^{({\rm AdS})}$ as described in section 2. 
In the second one, we determine the form of $H_{123}^{({\rm AdS})}$ 
by demanding that 
it satisfy the super-algebra, 
respect the $U(1)_Y$ symmetry of IIB supergravity, and 
match the data of the first subsection according to (\ref{hmap}).

\subsection{\label{sgdata} `Experimental' data} 

\begin{figure}
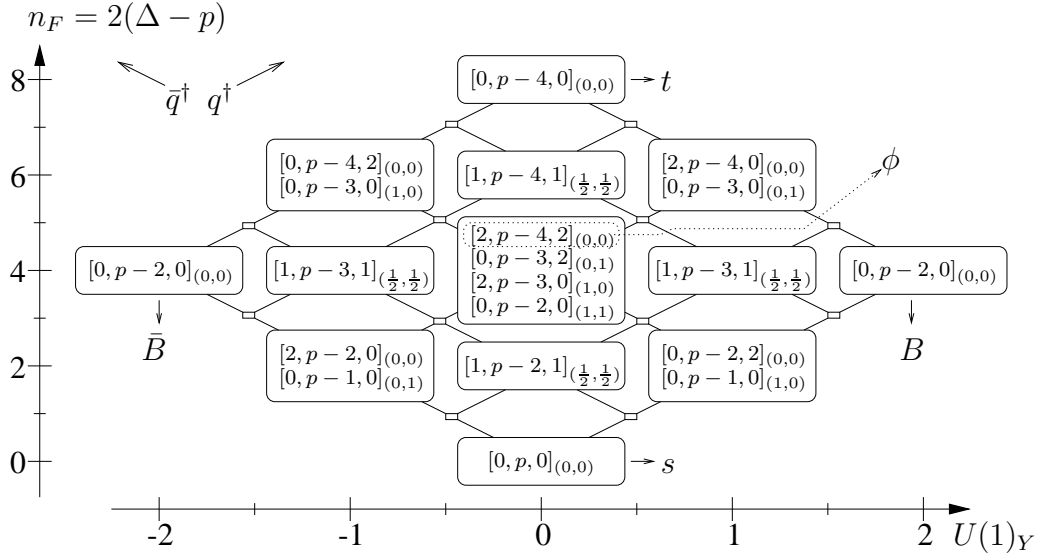

\begin{center}
\input diamond.pstex_t
\end{center}
\caption{Supergravity multiplet. Fermions are hidden in the small boxes}
\end{figure}

We begin by reviewing the structure of the IIB supergravity multiplet 
in AdS$_5 \times S^5$ \cite{Gunaydin:1984fk,Kim:1985ez}. 
After Kaluza-Klein reduction, 
the supergravity modes form a series of half-BPS multiplets 
of the $su(2,2|4)$ super-algebra.
Each multiplet is labeled by an integer $p$. 
Figure~2 shows how such a multiplet splits into several representations of 
the bosonic $so(2,4)\oplus so(6)$ bosonic algebra. 
The notation $[a,b,c]_{(i,j)}$ denotes the Dynkin label under $SO(6)$ 
and the $SU(2)\times SU(2)\approx SO(4) \subset SO(2,4)$ quantum number.
The AdS energy $\D$ of the ground state of a given supergravity mode
is the integer $p$ plus half of the number $n_F$ of supercharges 
needed to reach the given state from the ground state. 

In addition to the super-algebra quantum numbers, each mode is assigned 
the so-called $U(1)_Y$ charge. This $U(1)$ is the subgroup of 
the $SL(2,\IR)$ of the IIB supergravity, preserved by the AdS$_5\times S^5$ 
background. The dilaton-axion scalars form a complex scalar field 
with charge $\pm 2$ and combinations of NSNS and RR two-form fields have 
charges $\pm 1$, while the graviton and RR 4-form fields are neutral. 

Note that this $U(1)_Y$ is an exact symmetry 
of the AdS and pp-wave supergravity. In particular, the $U(1)_Y$ charge 
should be conserved in a cubic interaction involving three supergravity states.
Even in the full string theory in which the $U(1)_Y$ is broken, 
the selection rule will continue to hold when all three external states 
are supergravity states~\cite{Intriligator:1998ig,Intriligator:1999ff}.
This selection rule will play an important role 
in constructing the holographic cubic vertex in the pp-wave.

In the following, we present some explicit examples of 
the matrix element $H_{123}^{AdS}$ in (\ref{hmap}). 
They will impose severe constraints on $H_{123}^{PP}$ through (\ref{hmap}). 
For simplicity, we consider only scalar fields in $AdS_5$. 
There are four scalar fields that are also scalar on the $S^5$, 
as shown in Figure~2. The $s$ and $t$ fields are particular 
combinations of some components of the graviton and RR 4-form field. 
The complex $B$ field is the dilaton-axion pair which 
is related to the standard form $\tau = \chi + i e^{-\phi}$ by 
the conformal mapping,  
\be
B = \frac{\tau - \tau_0}{\tau + \tau_0}~,
\ee
so that, for any constant background value $\tau_0$, 
the $U(1)_Y$ symmetry acts {\it linearly} on $B$. 
The selection rule becomes manifest in this variable. 
Finally, we will also consider the field $\phi$ which 
is basically the graviton with both indices along the $S^5$ direction. 
So, it is a scalar in $AdS_5$, but a symmetric, traceless tensor on the $S^5$. 

The $S^5$ scalars $s, t, B$ transforms in $[0,k,0]$ representation of $SO(6)$. 
This $k$ is identified with the quadratic Casimir for spherical harmonics 
on $S^5$: 
$\del^2 Y = - k (k+4) Y$. As shown in Figure 2, 
$k$ is related to $\D$ and $p$ as $k = p - n_F/2 = \D - n_F$. 
More precise definition of the fields and their cubic couplings are 
summarized in the appendix~A and references therein.

\subsubsection*{Bosonic impurities}

The first class of amplitudes we consider involve three $s$-states. 
In the pp-wave set-up where a $U(1)$ R-charge is singled out, 
an $SO(6)$ representation splits into different $SO(4)$ representations. 
They correspond to the following operators in Yang-Mills, usually 
called the `scalar-impurity' operators in the pp-wave literature. 
\be
O_0 = \Tr( Z^J ), \;\;\;\;\;    
O_1 = \Tr( \phi Z^J ), \;\;\;\;\; 
O_2 = \sum_{l=0}^{J} \Tr( \phi Z^l \psi Z^{J-l} )~.
\ee
The number of impurities $n_B$ satisfies the relation $J +n_B = k = \D$. 
Since $J$ is conserved, 
$\D_{123}=k_{123}=(n_B)_{123}$ holds.
The following table summarizes several amplitudes. 
The numbers on the first column denote the 
number of impurities of each operator. The second column 
contains the value of LHS of (\ref{hmap}) normalized by 
$C^{(0)}_{123} \equiv \sqrt{J_1J_2J_3}/N$. The third and 
fourth columns contain the two factors on the RHS of (\ref{hmap}).
For later convenience, we define 
$q_i \equiv \sqrt{J_i/J_3}$ $(i=1,2)$.
\begin{equation}
\label{sdata}
\begin{array}{c|c|c|c}
(s^1 s^2 | s^3) & H_{123}/C_{123}^{(0)} \equiv V_s \D_{123}
& \frac{\D_{123}}{(\D_{123}/2)!} 
\left(\frac{J_1 J_2}{J_3} \right)^{\frac{\D_{123}}{2}}
& C_{123}/C_{123}^{(0)} \\ \hline 
(00|0) & 1 \cdot k_{123} & \D_{123} & 1 \\ 
(01|1) & q_2 k_{123} & \D_{123} & q_2 \\ 
(02|2) & q_2^2 k_{123} & \D_{123} & q_2^2 \\ 
(11|2) & q_1q_2 k_{123} & \D_{123} & q_1q_2 \\ 
(11|0) & q_1q_2 k_{123} & \frac{J_1J_2}{J_3} \D_{123} 
& \frac{1}{\sqrt{J_1J_2}} \\ 
(12|1) & q_1q_2^2 k_{123} & \frac{J_1J_2}{J_3}\D_{123} 
& \frac{1}{\sqrt{J_1J_3}} \\ 
(22|0) & q_1^2q_2^2 k_{123} 
& \left(\frac{J_1J_2}{J_3}\right)^2 \frac{\D_{123}}{2} & \frac{2}{J_1J_2}  
\end{array}
\end{equation}
The variable $k_{123}$ counts the impurity number violation,  
so it has definite integer values. However, we formally write it 
as if it is an undetermined variable, even when it vanishes, 
to facilitate comparison with the pp-wave vertex.

{}From the point of view of Kaluza-Klein reduction, 
operators with different scalar impurity configurations correspond to 
different spherical harmonics wave-functions on the $S^5$. 
The factors $V_s$ in the first column come from the spherical 
harmonics overlap integrals. As such, they are common for 
all supergravity fields that are scalar on the $S^5$. 
So, we will not separately discuss the effects of scalar impurities 
when we discuss other scalar fields than $s$ below. 

Next, we discuss the effect of the `vector impurities.' 
In the supergravity sector, the vector impurities are simply 
total derivatives acting on a given primary operator, 
\be
O = \Tr(Z, \phi, {\rm etc.}), \;\;\;\; 
O^{(1)}_\m = \partial_\m O, \;\;\;\;\; 
O^{(2)}_{\m\n} = \partial_\m \partial_\n O~. 
\ee
The resulting operators are descendants of the primary operator. 
In any CFT, correlators of descendants are completely 
determined by those of primaries. 
This fact is reflected in the supergravity computation. 
Primary state and descendant states are different wave functions 
of a same supergravity field. So, they share the same coupling constant. 
The only difference in $H_{123}$ then comes from the overlap integral 
of the three wave functions. The following table summarizes an 
explicit example of the $s$ field.
\begin{equation}
\label{vdata}
\begin{array}{c|c}
(s^1 s^2 | s^3) & H_{123}/C_{123}^{(0)} \equiv V_v V_s k_{123}
\\ \hline 
(00|0) & V_s \cdot k_{123}  \\ 
(01|1) & q_2 V_s k_{123} \\ 
(11|0) & q_1q_2 V_s k_{123} 
\end{array}
\end{equation}
The first row contains implicitly the entire table (\ref{sdata}) with 
no vector impurities. As one adds vector impurities, the wave function 
effect shows up as written in the last two rows. Note that the 
scalar impurities and vector impurities commute with each other. 
Note also that the `dynamic' part, $k_{123} = (n_B)_{123}$,  
counts only the scalar impurities but not the vector impurities. 
We see that the accidental $\IZ_2$ symmetry of the pp-wave string theory 
is broken. This will be crucial in determining the cubic vertex 
in the next subsection.

\subsubsection*{Fermionic impurities}

So far, we considered only the $s$ field which lies at the bottom 
of Figure 2. As we will see in the next subsection, it turns out 
that the amplitudes listed above are already sufficient to determine 
the supergravity vertex. Still, by comparing some amplitudes 
involving other states in Figure 2, we could verify in more detail, 
the validity of the vertex and as well as the duality map (\ref{hmap}).

For later convenience, we list all amplitudes involving $s$, $t$ and $B$ 
together. In the following table, it is understood that $V_v V_s$ 
is multiplied to each amplitude when bosonic impurities are added, 
and that $k_{123}$ counts only the scalar impurities. 

\hskip 1cm
\parbox[t]{6cm}{
\begin{displaymath}
\begin{array}{cc}
{\rm process} & H_{123}/C_{123}^{(0)} \\
(s^1s^2|s^3) & k_{123} \\
(t^1s^2|s^3) & q_2^8 (k_{123}+4) \\
(s^1s^2|t^3) & \CO(1/J^4) \\
(s^1t^2|t^3) & q_2^8 k_{123} \\
(t^1t^2|s^3) & \CO(1/J^4) \\
(t^1t^2|t^3) & (k_{123}+4) 
\end{array}
\end{displaymath}
}
\hskip 0.1cm
\parbox[t]{7cm}{
\begin{equation}
\label{tdata}
\begin{array}{cc}
{\rm process} & H_{123}/C_{123}^{(0)} \\
(s^1B^2|\bar{B}^3) & q_2^4 k_{123}  \\
(B^1\bar{B}^2|s^3) & q_1^4q_2^4 (k_{123}+4)  \\
(t^1B^2|\bar{B}^3) & q_2^4 (k_{123}+4) \\
(B^1\bar{B}^2|t^3) & q_1^4 q_2^4 k_{123} \\
 & \\
 &
\end{array}
\end{equation}
}

\noindent
The amplitudes are proportional to either $k_{123}$ or $(k_{123}+4)$.
Note that for each process $(12|3)$ listed in (\ref{tdata}), 
the constant shift to $k_{123}$ is always equal to $(n_F/2)_{123}$. 
In other words, including the shift, the coupling is 
proportional to $(n_B + n_F/2)_{123}$. 
This expression is most suitable for comparison with the 
pp-wave vertex. 
Alternatively, one can use $\D = k + n_F$  
to write the couplings as $(\D - n_F/2)_{123}$. 
Note that $(\D-n_F/2)$ is nothing but the scaling dimension 
of the chiral primary in the super-multiplet containing 
the given field.
This is natural since the coupling for a chiral primary and 
those for its super-descendants are expected to be proportional 
to each other. 

Finally, we compute amplitudes involving one $\phi$ field and 
two $s$ fields. Unlike the examples discussed above, 
this amplitude does not contain an explicit factor of 
$k_{123}$, because $\phi$ is not a scalar on the $S^5$. 

\hskip 1cm
\parbox[t]{6cm}{
\begin{displaymath}
\begin{array}{cc} 
(\phi^1 s^2 | s^3) & H_{123}/C_{123}^{(0)} \\
(\cdot 1| 1) & 0 \\
(\cdot 2| 0) & q_1^2q_2^4 \\
(\cdot 0| 2) & 0
\end{array}
\end{displaymath}
}
%\hskip 1cm
\parbox[t]{6cm}{
\begin{equation}
\label{phidata}
\begin{array}{cc} 
(s^1 s^2 | \phi^3) & H_{123}/C_{123}^{(0)} \\
(11| \cdot) & \CO(1/J^3) \\
(02| \cdot) & \CO(1/J^2) \\
 &
\end{array}
\end{equation}
}

\subsection{Construction of the vertex} 

We will closely follow the standard process of constructing the vertex, 
and our result will share many features with the previous proposals. 
However, compatibility with the duality map (\ref{hmap}) and 
the supergravity data listed in the previous subsection 
will lead to a final result different from all of the previous proposals. 

%%R
Let us briefly sketch the standard process.(See, for
example,~\cite{Green:1984fu,Pankiewicz:2003pg}). Quantization of the
string theory in the pp-wave is performed in the light-cone gauge. In
the light-cone gauge, space-time symmetries are implemented in the
interacting theory in two different ways.  All the generators that
leave the light-cone gauge fixing invariant are called
kinematical. They do not receive correction from the interactions and
can be promoted to local symmetries on the world-sheet. The remaining
generators are called dynamical and do receive corrections when the
interactions are turned on.  For the string theory in the pp-wave, the
light-Hamiltonian and a half of the 32 supercharges are the only
dynamical generators.

In principle, the cubic interaction part of the dynamical generators 
(Hamiltonian $H_3$ and super-charges $Q_3$ ) can be written as an 
operators in the string Fock space which change the number of strings. 
In practice, it is more convenient to translate $H_3, Q_3$   
into states $\ket{H_3}, \ket{Q_3}$ in the three string Hilbert space. 
Construction of $\ket{H_3}, \ket{Q_3}$ takes two steps. 
First, one builds a kinematical vertex $\ket{V}$ which manifestly respects 
all the kinematical symmetries. Then, the dynamical generators take the form
$\ket{H_3} = \hat{h}_3 \ket{V}$, $\ket{Q_3} = \hat{q}_3 \ket{V}$
The prefactors $\hat{h}_3, \hat{q}_3$ chosen such that 
the kinematical constraints are not spoiled and at the same time 
the commutation relation among the dynamical generators are also satisfied. 

\subsubsection*{Free theory: Review} 

The IIB supergravity in the pp-wave has manifest $SO(4)\times SO(4)$ 
rotation symmetry inherited from the $SO(2,4)\times SO(6)$ symmetry 
of $AdS_5\times S^5$. Following \cite{Pankiewicz:2003kj}, 
we use the vector index $i$ and bi-spinor indices $\ao, \aod$ 
for the first $SO(4)\subset SO(2,4)$, 
and ($i';\at, \atd$) for the second $SO(4)\subset SO(6)$. 
The Hilbert space of the free supergravity in the pp-wave 
is described by 8 bosonic oscillators 
$\{ a^i, (a^i)^\dagger ; a^{i'}, (a^{i'})^\dagger\}$ 
and 8 fermionic oscillators 
$\{ b_{\ao\at}, (b^\dagger)^{\ao\at} ; 
b_{\aod\atd}, (b^\dagger)^{\aod\atd} \}$.
The bosonic oscillators build up $(\Delta ; i, j )$ representation 
of $SO(2,4)$ and $[a,b,c]$ representation of $SO(6)$ in Fig.~1. 

The fermionic oscillators are identified with the kinematical super-charges 
up to a light-cone-momentum dependent factor 
(we assume $\a \equiv \a' p^+ > 0$ throughout this subsection), 
\be
q_{\ao\at} = \sqrt{\a} b_{\ao\at}, && 
(q^\dagger)^{\ao\at} = \sqrt{\a} (b^\dagger)^{\ao\at},  \xx
\bar{q}_{\aod\atd} = \sqrt{\a} b_{\aod\atd}, && 
(\bar{q}^\dagger)^{\aod\atd} = \sqrt{\a} (b^\dagger)^{\aod\atd}~. 
\ee
They form a super-multiplet of the same diamond shape as in Figure~1. 
The other 16 super-charges are dynamical. Explicitly, they are given by 
\be
Q^{\aod}{}_{\at} = (a^\dagger)^{\aod\ao}b_{\ao\at} 
- a_{\at\atd} (b^\dagger)^{\aod\atd}, &&
(Q^\dagger)_{\aod}{}^{\at} = a_{\ao\aod}(b^\dagger)^{\ao\at} 
- (a^\dagger)^{\atd\at} b_{\aod\atd}, \xx
\bar{Q}^{\ao}{}_{\atd} = (a^\dagger)^{\aod\ao}b_{\aod\atd} 
+ a_{\at\atd} (b^\dagger)^{\ao\at}, &&
(\bar{Q}^\dagger)_{\ao}{}^{\atd} = a_{\ao\aod}(b^\dagger)^{\aod\atd} 
+ (a^\dagger)^{\atd\at} b_{\ao\at}~. 
\ee
Note that all of them annihilate the oscillator vacuum, 
and that they are eigenstates of the $U(1)_Y$ symmetry.  
In what follows, the anti-commutators among the dynamical supercharges 
are important. The only nonvanishing terms are
\be
\{ Q^{\aod}{}_{\at} , (Q^\dagger)_{\bod}{}^{\bt} \} &=& 
2 \d^{\aod}_{\bod} \d^{\bt}_{\at} H + {\rm  rotations}, \xx
\{ \bar{Q}^{\ao}{}_{\atd} , (\bar{Q}^\dagger)_{\bo}{}^{\btd} \} &=& 
2 \d^{\ao}_{\bo} \d^{\btd}_{\atd} H + {\rm  rotations}~.
\ee

\subsubsection*{Cubic vertex} 

As a starting point, we use the kinematical vertex proposed
in~\cite{Chu:2002eu},
\be\label{sugraK}
\ket{V} = 
\exp{\left\{\frac{1}{2}  \sum_{r,s=1}^3 a^\dagger_{(r)}
M^{rs} a^{\dagger}_{(s)} \right\}}
\exp{\left\{- \sum_{i=1}^2 b^{\dagger}_{(i)}
q_i  
b^{\dagger}_{(3)} \right\}} \ket{v}_{123},
\ee
where $M^{rs}$ are the supergravity Neumann coefficients 
($q_i = \sqrt{|\a_i/\a_3|}$, $i=1,2$, as before),  
\be
\label{neu}
M = \pmatrix{ q_2^2 & -q_1q_2 & -q_1 \cr -q_1q_2 & q_1^2 & -q_2 
\cr -q_1 & -q_2 & 0 } .
\ee
In order for the prefactor not to spoil the kinematical constraint, it
should consist of the following combinations of oscillators.
\begin{equation}
\begin{array}{ll}
K^i = q_1 (a^\dagger_2)^i - q_2 (a^\dagger_1)^i , &
Y^{\ao\at} = q_1 (b_2^\dagger)^{\ao\at} - q_2 (b_1^\dagger)^{\ao\at}, \\
L^{i'} = q_1 (a^\dagger_2)^{i'} - q_2 (a^\dagger_1)^{i'} , &
Z^{\aod\atd} = q_1 (b_2^\dagger)^{\aod\atd} - q_2 (b_1^\dagger)^{\aod\atd}.
\end{array}
\end{equation}
Following the literature on the construction of the vertex, we assume that 
the prefactor has at most two powers of bosonic oscillators. 
It will be justified by matching the amplitudes via the duality map. 
The $U(1)_Y$ symmetry demands that 
$\ket{H_3}$ contains only terms with the same number of $Y$ and $Z$, 
and that the supercharges have terms like $Y^n Z^{n\pm 1}$ 
depending on their $U(1)_Y$ charges.
It is straightforward to enumerate all possible terms allowed to appear 
in the supercharges. Schematically, 
\be
\label{sch}
\ket{Q_3} &=& 
(c_1 LZ + c_2 K Y Z^2 + c_3 L Y^2 Z^3 + c_4 K Y^3 Z^4 ) 
\ket{V},
\xx
\ket{Q^\dagger_3}  &=& 
(d_1 K Y + d_2 L Y^2 Z + d_3 K Y^3 Z^2 + d_4 L Y^4 Z^3)
\ket{V},
\xx
\ket{\bar{Q}_3} &=& 
(\bar{c}_1 LY + \bar{c}_2 K Y^2 Z + \bar{c}_3 L Y^3 Z^2 
+ \bar{c}_4 K Y^4 Z^3 ) 
\ket{V},  
\xx
\ket{\bar{Q}^\dagger_3} &=& 
(\bar{d}_1 K Z + \bar{d}_2 L Y Z^2 + 
\bar{d}_3 K Y^2 Z^3 + \bar{d}_4 L Y^3 Z^4)
\ket{V}~.
\ee
Define the sum of super-charges in the free theory,  
\be
Q = \sum_{r=1}^{3} Q_r^{(2)}~,
\ee
for the four kinds of dynamical supercharges. The super-algebra 
at the cubic level demands that 
\be
Q \ket{Q^\dagger_3} + Q^\dagger \ket{Q_3} = \ket{H_3}, \;\;\; 
\bar{Q} \ket{\bar{Q}^\dagger_3} + \bar{Q}^\dagger 
\ket{\bar{Q}_3} = \ket{H_3}~, 
\ee
and that similar equations with ${\rm RHS} =0$ hold 
for all the other combinations of super-charges.

A straightforward but tedious calculation gives a seemingly over-constrained 
set of linear equations among the coefficients in (\ref{sch}).
However, it turns out that many of the relations are linearly dependent, 
and there are three independent solutions. 
The solution for the cubic Hamiltonian is given by,
\be
\ket{H_3}  &=& 
h_0 \Big( (L^2 - K^2)(1+Y^4Z^4) + 2
KL (YZ +Y^3 Z^3)- K^2 (YZ)^2 + L^2 (YZ)^2 \Big) \ket{V} 
\xx
&&
+ h_- (K^2 + L^2) \ket{V}
+ h_+ (K^2+ L^2 +8) (YZ)^4 \ket{V} , 
\ee
where $h_0, h_-,h_+$ are so far undetermined constants. 
As expected, the super-algebra alone does not fix the vertex completely.
It is now time to use our knowledge on holography discussed above. 
Using the conservation laws for the kinematical symmetries and 
$q_1^2 + q_2^2 = 1$, one can show that 
\be
(L^{i'})^2 \ket{V} = (n_B)_{123} \ket{V} ,
\ee
that is, $(L^{i'})^2$ counts the change in the number of scalar impurities. 
Similarly, one can show that the $(K^i)^2$ term counts vector impurities. 
However, we saw that vector impurities contribute only the 
`wave-function factor' $V_v$ and do not affect the coupling constant. 
This fact demands via the duality map (\ref{hmap}) 
that $\ket{H_3}$ should not contain a factor of $(K^i)^2$ 
when all external states are $SO(4)\times SO(4)$ scalars. 
This implies that $h_- = h_0 = h_+$.  
The overall normalization can be fixed by matching any one of 
the non-zero amplitudes listed in (\ref{sdata}). 
All in all, the final answer is\footnote{
Our definitions for the products of $Y$ and $Z$ are slightly 
different from those of~\cite{Pankiewicz:2003kj}: 
$Y^2 = Y^2_P/2, Y^3 = Y^3_P/3, Y^4 = Y^4_P/12$ and similarly for $Z$.
Accordingly, the polynomials $v$ and $s$ appearing in \eq{svp} 
should be understood as $s(Y) = Y + i Y^3$,
$v^{ij}= \d^{ij}(1+Y^4)(1+Z^4) 
- i[(Y^2)^{ij}(1+Z^4)-(Z^2)^{ij}(1+Y^4)]+(Y^2Z^2)^{ij}$ and 
$v^{i'j'}= \d^{i'j'}(1-Y^4)(1-Z^4) 
- i[(Y^2)^{i'j'}(1-Z^4)-(Z^2)^{i'j'}(1-Y^4)]+(Y^2Z^2)^{i'j'}$
.
},
\be
\label{sganswer}
\displaystyle
\ket{H_3} &=& C^{(0)}_{123}  \Big( (L^{i'})^2+ 
\{ (L^{i'})^2+4 \} Y^4 Z^4 
+ K^{\aod\ao}L^{\atd\at} (Y_{\ao\at}Z_{\aod\atd} 
+ Y^3_{\ao\at} Z^3_{\aod\atd}) \Big) \ket{V}
\xx
&& + \frac{C^{(0)}_{123}}{2}  \Big( 
L^{\atd\at}L^{\btd\bt} Y^2_{\at\bt}Z^2_{\atd\btd} - 
K^{\aod\ao}K^{\bod\bo} Y^2_{\ao\bo}Z^2_{\aod\bod}  
\Big) \ket{V}~.
\ee
Note that we used only the `bosonic impurity' part of the previous 
subsection to determine the vertex. The $(L^{i'})^2$ factor gives $k_{123}$ 
part of (\ref{sdata}) and (\ref{vdata}). The wave-function factor
$V_s$ and $V_v$ match exactly elements of the bosonic Neumann matrix. 
 
Now, all the amplitudes containing `fermionic impurities' 
provide further checks on the vertex. 
First, note that the factor $((L^{i'})^2+4)$ multiplying $Y^4Z^4$ matches
$(k_{123} +4)$ in (\ref{tdata}). 
In fact, the $(YZ)^0$ term and $(YZ)^4$ terms can be combined into 
$(n_B + n_F/2) (1+Y^4Z^4)\ket{V}$.  
The factors of $q_1, q_2$ in (\ref{tdata}) come from both 
the fermionic Neumann matrix, and for those proportional to 
$(k_{123}+4)$, also from $Y^4Z^4$. 
Finally, one can check that the $L^2(YZ)^2$ term gives 
the non-vanishing entry in the table (\ref{phidata}) 
for the $(\phi s s)$ amplitudes. 

\section{String theory vertex}

We now turn to the task of generalizing the supergravity
vertex~\eq{sganswer} to the full string theory. We first need to
enlarge the Fock space to include also the states created by the
stringy oscillators $a^\dagger_n$ and $b^\dagger_n$. Then we need to
find a $3$-string vertex satisfying two main constraints: it should
realize the pp-wave super-algebra at cubic level and it should reduce
to the supergravity expression~\eq{sganswer} in the $\mu\alpha_i\to 0$
limit. In principle one could proceed in a systematic way as done in
the previous section for the BPS sector, but this exhaustive approach
is rather complicated at the string level. It is easier and also more
instructive to derive the cubic vertex by combining the results
derived in previous works.  

We start by choosing a coherent state that realizes the kinematical
part of the algebra $\ket{V} = E_a\, E_b\, \ket{v}_{123}$, where the
two terms contain the bosonic and the fermionic contributions
respectively. Let us recall their explicit expressions\footnote{For the
conventions on the string oscillators and the explicit definition of
the Neumann matrices the reader is referred
to~\cite{Pankiewicz:2002gs} and references therein.}.  The bosonic
exponential reads~\cite{Spradlin:2002ar}
\be
E_a = \exp{\left\{\frac{1}{2}  \sum_{r,s=1}^3 a^\dagger_{n (r)}
N^{rs}_{nm} a^{\dagger}_{m (s)} \right\}}~.
\ee
The string Neumann coefficients are usually written in terms of
products of infinite matrices. From this formal definition many
properties can be derived~\cite{Schwarz:2002bc,Pankiewicz:2002gs},
however it is difficult to obtain an explicit value of the
$N^{rs}_{nm}$ in terms of $n,m$ and the $\a_i$'s, since the original
product expression contains the inverse of an infinite matrix. A
detailed study of the Neumann coefficients for $\mu\not = 0$ can be
found in~\cite{He:2002zu,Lucietti:2003ki,Lucietti:2004wy}.
In the fermionic sector we will use the coherent state introduced
in~\cite{Chu:2002wj}, which can be written in the  $SO(4)\times SO(4)$
notation as done in~\cite{Pankiewicz:2003kj} 
\be
E_b = \exp\left[\sum_{r,s=1}^3\sum_{m,n\geq 0}
\Big(b^{\a_1\a_2\,\dagger}_{-m(r)}b^{\dagger}_{n(s)\,\a_1\a_2}+
b^{\dot\a_1\dot\a_2\,\dagger}_{m (r)}
b^{\dagger}_{-n(s)\,\dot\a_1\dot\a_2}\Big)
Q^{rs}_{mn}\right]
~.
\ee

As we reviewed in the introduction, this kinematical part can be
completed into a fully supersymmetric interacting Hamiltonian in (at
least) two physically different ways. One possible completion was
first
proposed by~\cite{Spradlin:2002ar,Spradlin:2002rv,Pankiewicz:2002tg} in
the $SO(8)$ formalism. Subsequently the same vertex was recast in
the $SO(4)\times SO(4)$
language~\cite{Pankiewicz:2003kj,Pankiewicz:2003ap} and here we will
stick to the $SO(4)\times SO(4)$ notation 
\be
\ket{H_3}_I  & = & -\; 
\Bigg[\Big(K_i\tilde{K}_j+\frac12\, \delta_{ij}\Big) v^{ij} (Y,Z)
-\Big(L_{i'}\tilde{L}_{j'} + \frac 12\, \delta_{i'j'}\Big) v^{i'j'} (Y,Z)
\label{svp} \\ \nonumber & - & 
K^{\dot\a_1\a_1} \tilde{L}^{\dot\a_2\a_2} s_{\a_1\a_2}(Y)
s^*_{\dot\a_1\dot\a_2}(Z) - \tilde{K}^{\dot\a_1\a_1} L^{\dot\a_2\a_2}
s^*_{\a_1\a_2}(Y) s_{\dot\a_1\dot\a_2}(Z) \Bigg]\ket{V}~,
\ee
where again we follow the conventions and notation
of~\cite{Pankiewicz:2003kj}, except for the normalization the bosonic
constituents which is slightly different, 
\be
K^i = \sqrt{\frac{\a'}{2 \mu |\a_1 \a_2 \a_3|}}
\; K_{P}^i 
~~,~~~
L^{i'} = \sqrt{\frac{\a'}{2 \mu |\a_1 \a_2 \a_3|}}
\; K_{P}^{i'}~.
\ee

Another possibility for writing a supersymmetric vertex is discussed
in~\cite{DiVecchia:2003yp}, where it was proposed to use simply the
free Hamiltonian as prefactor of the coherent state
\be\label{even}
\ket{H_3}_{D} = \sum_{r=1}^3 H_r \ket{V}~.
\ee
However, it was first noticed in~\cite{Dobashi:2004nm} that neither of
the two vertices~\eq{svp} and~\eq{even} have the expected behaviour
from the holographic point of view. We can rephrase this observation
in a somehow different way by using the results of the previous
section: the supergravity limit of the vertices~\eq{svp} and~\eq{even}
breaks the relation~\eq{inth} because they contain some $K_0^2$ term
in the prefactor which is absent on the AdS side. 
So it was proposed~\cite{Dobashi:2004nm} that the holographic cubic
vertex for the pp-wave background is proportional to $\ket{H}_I +
\ket{H}_{D}$. It is interesting to notice that this combination
reproduces, when restricted to the scalar bosonic oscillators, the
`phenomenological' prefactor introduced in~\cite{Chu:2003qd} to explain
the field theory results from a string theory point of view.

However, a closer comparison between the proposed vertex $\ket{H}_I +
\ket{H}_{D}$ and the large $J$ limit of the AdS couplings shows that
relation~\eq{inth} is not yet satisfied. In fact, when we restrict the
combination $\ket{H}_I + \ket{H}_{D}$ to the supergravity sector, the
only term that perfectly matches the AdS expectation is the one
without fermionic insertions (of $Y_0$ and $Z_0$). However, it is not
difficult to see how to modify the vertex~\eq{even} in such a way that
its combination with~\eq{svp} gives the expected supergravity answer.
First we should add two pieces quartic in the fermions ($Y^4$ and
$Z^4$) so that at the supergravity level the $U(1)_Y$ violating terms
of~\eq{svp} are canceled. Then a contribution with eight fermionic
insertions ($Y^4 Z^4$) should be added to match the second term
in~\eq{sganswer}. Thus our final proposal for 
the holographic cubic vertex is
\be\label{holov}
\ket{H} = \frac{C^{(0)}_{123} }{2}\; \Big( \ket{H}_I + \ket{H}_{II} \Big)~,
\ee
where
\be\label{geven}
\ket{H_3}_{II} = \left( \sum_{r=1}^3 H_r \right)
\Big( 1+Y^4 \Big) \Big( 1+Z^4 \Big)
\ket{V}~.
\ee
Clearly this contribution to the vertex is a natural generalization of
the~\eq{even} and it satisfies by itself the supersymmetry
constraints. In fact the combination $(1+Y^4+Z^4+Y^4 Z^4 )\ket{V}$
satisfies all the requirement related to the kinematical part of the
pp-wave algebra. Thus it can be `dressed' with the free supercharges
or Hamiltonian as done in~\cite{DiVecchia:2003yp} in order to produce
a consistent system of interacting correction to the free
generators. Notice also that the commutation of the kinematical
constraints with $\sum_r H_r$ is again a combination of the kinematical
constraints and thus does not spoil the properties of the coherent state 
$\ket{V}$.

\subsection{Some checks on the string vertex}

From the gauge theory point of view the holographic vertex contains a
great deal of information on non-BPS quantities, since the dependence
of the Neumann matrices on $\mu\alpha_i$ in~\eq{holov} translates, in
the SYM theory, into the exact dependence on the 't~Hooft
coupling. Moreover in the non-supersymmetric sector, the comparison
with the gauge theory is the only way at our disposal to check the
correctness of the proposal~\eq{holov}.  However, it is still not
entirely clear how to relate in general string and gauge theory
results, since the dictionary~\eq{hmap} between $3$-point correlators
in the two descriptions has been derived only in the supergravity
approximation. The authors of~\cite{Dobashi:2004nm} proposed a small
modification of~\eq{hmap}
\be \label{dic2}
\D_{123} \left(\frac{J_1 J_2}{J_3} \right)^{\frac{\D_{123}}{2}}
C_{123} =
\left(f \right)^{-\frac{\D_{123}}{2}} 
{\G\left(\frac{\D_{123}}{2} +1\right)} \,H^{\rm (PP)}_{123} ~,
\ee
where $f$ is a combination that appears in various places of the
string computations (see, for instance, \cite{Pankiewicz:2002gs}): $f
= (1-4 \mu\a K)$. With this prescription the $3$-point functions among
BPS states are independent of $\mu\alpha_i$, even if the full string
vertex~\eq{holov} is used to compute the correlator. This can be
checked by using the relation between the stringy Neumann coefficients
and the supergravity ones: $N^{ij}_{00} = f M^{ij}$ for $1\leq
i,j\leq2$ and $N^{i3}_{00} = M^{i3}$. The requirement to have constant
$3$-point functions among BPS states is in accordance with the
expected non-renormalization theorem~\cite{Lee:1998bx} of the SYM
correlators among three BPS operators. Of course it would be very
interesting to {\em derive~}\eq{dic2} in order to check the
non-renormalization theorem, instead of imposing it. Moreover it is
quite likely that other $\alpha'$-dependent modifications will appear
in the exact dictionary between $ C_{123}$ and $H^{\rm (PP)}_{123}$.
However, if we focus on the first order in the $\lambda'$ expansion,
the simple Eq.~\eq{dic2} is able to capture completely the relation
between gauge theory and string theory. Let us briefly summarize the
evidence collected so far supporting this proposal.

-- The first thing we want to verify is that the new terms introduced
in~\eq{holov} do not spoil the agreement between string and gauge
theory correlators found in previous works. It is clear that for
purely bosonic amplitudes the new terms present in~\eq{geven} are
irrelevant and so all the checks already done in this
subsector\footnote{This applies also to the recent
  papers~\cite{Dobashi:2004ka}, as well as to previous
  works~\cite{Chu:2003qd,Georgiou:2003aa}.}  supports our
proposal~\eq{holov}. On the contrary, the amplitudes with four or more
fermionic impurities are sensitive to the novelties contained
in~\eq{holov}. However, in the situation studied
in~\cite{Dobashi:2004nm}, the four fermions are divided in an impurity
preserving way, that is two of them act on the ingoing state (the one
with negative $\a_i$) and the others act on the outgoing states (those
with $\a_i>0$). In this case, the new contributions in~\eq{geven}
appear only at the next-to-leading order in $\lambda'$, In fact the
$Y^4$ and $Z^4$ terms appearing in $\ket{H}_{II}$ are multiplied by
$\sum_r H_r$ and in the impurity preserving processes $\sum_r H_r \sim
O(\lambda'$). Similar terms quartic in the fermions are present also
in $\ket{H}_{I}$, but they do not have the energy difference as
additional factor and so their contribution survives also at the first
order in the $\lambda'$ expansion. Thus the $O(\lambda')$ result for
these amplitudes is again in agreement also with the
vertex~\eq{holov}. This situation is very similar to that encountered
in the study of the processes where the number of impurities is
preserved, but their flavor changes (like the process considered
in~\cite{Chu:2003qd}). Also in this case only $\ket{H}_I$ contributes
to the leading order result of the string amplitude.

-- In the truly non-impurity preserving processes, where
also the number of impurities changes from the operator $O_3$ to the
operators $O_1$ and $O_2$, the full vertex~\eq{holov} enter. We have
already seen in section~3 that the new terms in~\eq{geven} are
necessary to have agreement with the large $J$ limit of supergravity
results. In the BPS sector this ensures that the string amplitudes do
agree also with the gauge theory answer, thanks to the standard
AdS/CFT duality. Let us see how this works by focusing for instance on
the sixth case in the table~\eq{sdata}. The relevant operators are
\be
O_1 = \Tr(\phi Z^{J_1} )~, \;\;\;\;\;    
O_2 = \frac{1}{\sqrt{J_2}} \sum_{l=0}^{J_2} \Tr(\bar\phi Z^l \psi Z^{J_2-l} )
~, \;\;\;\;\; 
O_3 = \Tr( \psi Z^{J_3} )~,
\ee
and it is straightforward to see that the gauge theory combinatorics
reproduces in the large $J$ limit the third column of~\eq{sdata}
\be
\langle \bar O_3(x_3)\, O_2(x_2)\, O_1(x_1) \rangle = 
\frac{1}{\sqrt{J_1 J_3}} \frac{C^{(0)}_{123}}
{|x_1-x_2|^{2\b_3}|x_2-x_3|^{2\b_1}|x_3-x_1|^{2\b_2}}~.
\ee
On the string side one obtains 
\be
{}_{123} \bra{v}\; a_{0 (3)}^{\bar\psi} \,  
a_{0 (2)}^{\bar\phi} a_{0  (2)}^{\phi}  \,
a_{0 (1)}^{\phi} \ket{H} = 2 N_{00}^{12} N_{00}^{23} = 
2 f M_{00}^{12} M_{00}^{23} ~.
\ee
By using~\eq{neu} and, in this case, $\Delta_{123} =2$, we see that
this is equal to the first column of the table~\eq{sdata} multiplied
by the factor $f$ which is the difference between the supergravity and
the full Neumann matrices for the elements $N^{ij}_{00}$ with $1\leq
i,j\leq 2$. However, the dictionary~\eq{dic2} was engineered to cancel
the factors of $f$ and in fact we get the same $\mu$--independent
answer obtained in section~3. It is also easy to study the same
amplitude in the string case, where the second operator is replaced
by
\be\label{2io}
O_2 = \frac{1}{\sqrt{J_2}} \sum_{l=0}^{J_2} \Tr(\bar\phi Z^l \psi Z^{J_2-l} )
\;{\rm e}^{2\pi i \frac{n l}{J_2+1}}~.
\ee
In this case the tree-level result on the gauge theory side is zero,
because the phase forces the final sum over $l$ to vanish. On the
string side, the only difference with the BPS case is that now the
result is proportional to the Neumann matrices $N_{0n}^{12}
N^{23}_{n0}$, while before we had $n=0$. By using the results
of~\cite{He:2002zu}, we find that in this case the first non-trivial
contribution to the RHS of~\eq{dic2} starts at order $\lambda'$, in
agreement with the gauge theory results which fixes the tree-level
contribution to be zero. 

-- The last case of table~\eq{sdata} presents the prototypical case of
impurity non-preserving processes. In this case both `outgoing'
operators contain two impurities. On the gauge theory side the large
$J$ limit of this amplitude does not change when we pass from BPS
operators to stringy ones with the BMN phase (like that of
Eq.~\eq{2io}). This is because only particular terms in the sum
defining the operators contribute to the amplitude in the {\em planar}
approximation and ${\rm e}^{2\pi i n/J} \to 0$ for any $n\not= 0$ in
the BMN limit. On the string side this observation implies that the
elements $N^{ij}_{nm}$ with $1\leq i,j\leq 2$ of the Neumann matrices
are, {\em at leading order in} $\lambda'$, basically the same as the
zero-mode elements. Again by using the results of~\cite{He:2002zu} one
can check that this is indeed the case. Thus we can use the agreement
between string and supergravity/CFT results at the BPS level in order
to claim that impurity non-preserving amplitudes agree at leading
order in $\lambda'$ also for generic non-BPS states.

\section{Discussion}

In the usual approach to the BMN duality, one first tries to build the
pp-wave string Hamiltonian by using only the internal consistency of
the theory and then looks for a string/SYM dictionary compatible with
the string vertex. Since the two vertices~\eq{svp} and~\eq{even} are
rather different, they motivated two different ways to relate string
theory interactions with the dual gauge theory results. Inspired by
the string bit proposal~\cite{Verlinde:2002ig,Vaman:2002ka}, various
authors \cite{Gross:2002mh,Pearson:2002zs,Gomis:2002wi,Gomis:2003kj}
studied the relation between the string vertex~\eq{svp} and the mixing
between single and double trace operators on the field theory side
(see also~\cite{Georgiou:2003kt,Georgiou:2004ty} for further checks in
this direction). In particular, they proposed to identify the
$3$-string couplings derived from~\eq{svp} with the matrix elements of
the gauge theory dilatation generator in a particular basis in the
space of the single and double trace operators. On the other hand the
vertex~\eq{even} was motivated by realizing in string theory the
proposal of~\cite{Constable:2002hw} that relates the $3$-string
couplings with the correlators among the BMN operators on the gauge
theory side. Notice that also this point of view is consistent with
the string bit picture, since it identifies, in the $\mu\alpha_i
\to\infty$ limit, the world-sheet dynamics with the free contractions
among the constituents of the three operators (see for instance the
figures for the $3$-point function in~\cite{Constable:2002hw}
and~\cite{Russo:2004kr}). Even though these two proposals were checked
in various different cases, the situation was not completely
satisfactory. First the agreement between string and field theory
results was checked only at leading order in~$\lambda'$. Then, on the
conceptual ground, it was rather unclear the role played in the
duality by the gauge theory operators that are exact eigenstates of
the dilatation generator. At leading order in $g_2$ these eigenvectors
are a particular combination of single and double trace
operators. However, on the one hand the comparison between the string
vertex~\eq{even} and the gauge theory results gave agreement only by
using the original BMN
operators~\cite{Kiem:2002xn,Huang:2002wf,DiVecchia:2003yp} and
ignoring the multi-trace corrections of the dilation eigenvectors.  On
the other hand the string/gauge theory comparison with the
vertex~\eq{svp} required a mixing between single and double traces
that was {\em different} from the one necessary to define the
dilatation eigenvectors. In fact the field theory computations
of~\cite{Chu:2003qd,Georgiou:2003aa}, that are done with the
dilatation eigenvectors, represented for long time a puzzle from the
string point of view, since they seemed to be not related to either of
the two vertices~\eq{svp}-\eq{even}.

In order to overcome these problems, in this paper we reversed the
approach commonly adopted so far and constructed a $3$-string
interaction in the PP-wave background by taking into consideration
all possible information from different descriptions 
from the very beginning.
In particular, we study systematically the constraints on the string
dynamics coming from the large $J$ limit of AdS$_5 \times S^5$
supergravity. Our results confirm the physical picture of Dobashi and
Yoneya~\cite{Dobashi:2004nm} and show how the string vertex has to be
generalized in order to describe correctly also impurity
non-preserving processes. Moreover, as explained in section~5
of~\cite{Dobashi:2004nm}, this approach is able to explain also the
partial success, for impurity preserving processes, of the previous
string/gauge theory comparisons (see the discussion above).  For these
processes, it is possible to separate the contributions coming from
the free field theory combinatorics from those responsible of the
operator mixing and map them into the $\ket{H_3}_D$ and
$\ket{H_3}_{SV}$ parts of the full string vertex $\ket{H_3}$.

Let us conclude by summarizing here the main results derived in this
paper and focusing on the properties of the vertex~\eq{holov}.
A first unexpected feature is that the string interaction must break
the $\IZ_2$ symmetry of the pp-wave background, which, on the
contrary, was preserved by the free spectrum. It was first noticed
in~\cite{Chu:2002eu} that SV vertex~\cite{Spradlin:2002ar} had a
definite parity under this discrete symmetry. It was further proposed
that one should build a different $3$-string vertex, with opposite
parity, in order to make a direct comparison with gauge theory
correlators possible.  This idea was in striking contrast with the
belief that there was a unique possible interacting Hamiltonian
realizing the relevant supersymmetry algebra. However an explicit
realization of this proposal~\cite{DiVecchia:2003yp} showed the
necessity of further constraints in order to fix completely the string
cubic Hamiltonian. However, it turns out that the behaviour under the
$\IZ_2$ symmetry is not a reliable input for fixing the form of the
string vertex. A first signal that this $\IZ_2$ was not a good
symmetry at the interacting level came from the
study~\cite{Chu:2003ji} of field theory correlators among dilatation
eigenstates containing vector impurities. Here instead we used the
insights coming from supergravity and we showed that the interacting
Hamiltonian must contain both odd and even terms under $\IZ_2$.
Moreover the vertex~\eq{holov} singled out by our analysis contain new
$SO(4) \times SO(4)$ preserving combinations of the various building
blocks~\cite{Pankiewicz:2002gs,DiVecchia:2003yp}, realizing once more
a situation quite common in physics (i.e.  everything that is not
forbidden is compulsory). It is natural at this point to ask whether
it is necessary to add further corrections to Eq.~\eq{holov} that are
not captured by our supergravity analysis. Although this seems 
unlikely we can not rule out such corrections. For instance we still use as
an additional input the requirement that the prefactor is at most
quadratic in the bosonic oscillators. In order to clarify completely
this point it would be necessary to derive the cubic Hamiltonian from
first principles, for instance by applying a standard path integral
approach also in the derivation of the prefactor (and not only for the
exponential part, as it was done in~\cite{Russo:2004kr}).

Another interesting aspect of our string proposal is to see how the
$U(1)_Y$ symmetry is realized at the level of BPS (or supergravity)
interactions. Actually this is a general observation, not restricted to the
particular pp-wave background we are focusing on. In fact a similar
pattern appears also in the construction of the flat space IIB string
field theory: in~\cite{Green:1983tk} it was noticed that the $U(1)_Y$
symmetry forces the supergravity prefactor to be quartic in the
fermionic fields. However, the full string
construction~\cite{Green:1983hw} requires the presence of other terms
that survive also when the amplitudes are restricted to the
supergravity sector. The original observation in~\cite{Green:1983hw}
was that these new terms are proportional (at the supergravity level)
to the difference of the free Hamiltonians ($\sum_r H_r$) and thus are
zero on-shell. In order to have a conserved $U(1)_Y$ symmetry also
off-shell, \cite{Green:1983hw}~proposed that the $U(1)_Y$ generator
should get corrections in the interacting theory. Here we show that
there is a simpler way out: one can define the off-shell cubic
Hamiltonian for the flat space to be a simple combination of the
Brink, Green and Schwarz vertex and of the following vertex
\be\label{bgs-gen}
\ket{H_3} = \ket{H_3}_{BGS} - \left(\sum_r H_r\right) 
\left(1 + \prod_{a=1}^8 Y^a_{BGS} \right) \ket{V}_{BGS}~,
\ee
where we are now using the conventions
of~\cite{Green:1983hw}. Notice that the additional piece is irrelevant
if we just want to compute on-shell scattering amplitudes because in
flat space the energy is conserved. Thus previous checks on S-matrix
elements like those in~\cite{Hornfeck:1987wt} are not affected by the
modification proposed here. However, the inclusion of the new terms
in~\eq{bgs-gen} yields a $U(1)_Y$ preserving (supergravity) vertex
also off-shell. In the pp-wave case this feature is necessary since we
clearly do not want any conservation law on $H_r$ in the physical
observables and so the terms proportional to $\sum_r H_r$ can not be
disregarded. However, the modification proposed in~\eq{bgs-gen} is
important also in flat space every time one needs to go off-shell.
Problems of this type are constructing a $4$-string vertex by
sewing two $3$-string vertices or computing the energy of an
arbitrary string configuration including the cubic contributions $H_3$.
It is known that the vertex $\ket{H_3}_{BGS}$ is incomplete and can
not be used to deal consistently with these questions. Because of
these problems it has been proposed that the light-cone string field
theory contains also quartic
terms~\cite{Greensite:1986gv,Greensite:1987sm,Green:1987qu,Greensite:1987hm}. 
It would be very interesting to reconsider these issues by
using the $3$-string vertex~\eq{bgs-gen} to see whether it can
provide a different completion of $\ket{H_3}_{BGS}$ that does not
require quartic corrections.

Also on the field theory side the string vertex~\eq{holov} together
with the duality map~\eq{dic2} yields some interesting and
counterintuitive consequences. For instance, it is common to write the
BMN operators by focusing only on the leading term in the $J\to\infty$
limit, even if in principle they are combinations of various
contributions with the same quantum numbers\footnote{The importance of
  certain compensating terms, subleading in the $J\to\infty$ limit,
  was already stressed in~\cite{Beisert:2002tn,Chu:2003ji}.}. This is
the so-called `dilute gas approximation' where the impurities are
always thought to be far apart from each other. However in the
impurity non-preserving processes this approximation breaks down even
in the simplest situations, since in the holographic
dictionary~\eq{hmap} between gauge and string theory correlators there
is a compensating $J$-dependent factor. This term plays an important
r\^ole in the correlators with {\em different} barred and unbarred
operators (i.e. with different `ingoing' and `outgoing' states). In
the dilute gas approximation this kind of amplitudes is trivially
vanishing, while on the string side the corresponding processes are
non-zero, since they get a non-zero contribution from the various term
in the prefactor containing the fermionic insertions. The presence of
the compensating factor in~\eq{hmap} enhances the contributions coming
from the subleading (in $J$) terms in the definition of the BMN
operators and gives a non-zero answer also on the gauge theory side.

Finally a very important open issue is the full justification of the
holographic dictionary. For example, the duality map~\eq{dic2}, if
correct, can provide a resolution to the puzzle of fractional powers
of $\lambda'$ raised in~\cite{Klebanov:2002mp}. While the map in the
supergravity sector~\eq{hmap} has been derived from directly from the
rules of the AdS/CFT duality, its string generalization~\eq{dic2} has
been proposed~\cite{Dobashi:2004nm} by imposing the
non-renormalization of the $3$-point BPS correlators. It is clearly
important to test and possibly completely fix this holographic
dictionary. Two complementary approaches are possible: either one can
work from the bulk point of view and generalize the physical picture
sketched in section~2.3 from the particle to the string case, or one
starts from the field theory by pushing the computations to the
subleading order in $\lambda'$.

\vskip 1cm 

\centerline{\bf Acknowledgement}

\vskip .5cm

We are grateful to A. Tanzini for useful discussions.

\appendix

\section{Coupling constants}

We summarize the cubic couplings needed to compute the amplitudes 
in section 3. The $(sss)$ coupling was first computed in \cite{Lee:1998bx}.  
Couplings for two $s$ and another arbitrary field were worked out in 
\cite{Arutyunov:1999en,Lee:1999pj}. 
Other couplings listed below can be derived in a similar way. 
We follow the notations of \cite{Lee:1999pj}. 
The part of AdS$_5$ supergravity action relevant to our discussion 
can be written as 
\be
S = \frac{N^2}{8 \pi^2} \int d^5 x \sqrt{-g} \{ L_2 + L_3 \}~. 
\ee
The quadratic Lagrangian takes the following form 
\be
L_2 = - 
\sum_{\varphi = s, t, \phi} 
\frac{A_\varphi}{2} \{(\del \varphi)^2 + m_\varphi^2 \varphi^2 \} 
- A_B \{ |\del B|^2 + m_B^2 |B|^2 \}~.
\ee
The mass of each scalar is determined by the usual relation $m^2 = \D(\D-4)$ 
and the relation between $\D$ and $k$ mentioned in subsection 3.1. The 
normalization constants are given by 
\be
\textstyle A_s = 2^5 \frac{k(k-1)(k+2)}{k+1} z(k), \;\;
A_t = 2^5 \frac{(k+4)(k+5)(k+2)}{k+3} z(k) , \;\; 
A_B = z(k),\;\; A_\phi = \half z(k)~.
\ee
The cubic Lagrangian is given by
\be
L_3 &=& 
-\frac{1}{6} G^{(sss)}_{123}s^1s^2s^3 
-\frac{1}{6} G^{(ttt)}_{123}t^1t^2t^3 
-\frac{1}{2} G^{(tss)}_{123}t^1s^2s^3 
-\frac{1}{2} G^{(stt)}_{123}s^1t^2t^3 
\xx
&&-\frac{1}{2} G^{(\phi ss)}_{123}\phi^1s^2s^3 
- G^{(sB\bar{B})}_{123}s^1B^2\bar{B}^3 
-G^{(sB\bar{B})}_{123}t^1B^2\bar{B}^3 , 
\ee
where the coupling constants are given by
\be
(s^1s^2s^3) &:& 2^9 \frac{\a_1\a_2\a_3}{(k_1+1)(k_2+1)(k_3+1)} 
\frac{(\s+2)!}{(\s-3)!} a(k_1,k_2,k_3) \langle C^1 C^2 C^3 \rangle, \xx
(t^1t^2t^3) &:& 2^9 \frac{(\a_1+2)(\a_2+2)(\a_3+2)}{(k_1+3)(k_2+3)(k_3+3)}
\frac{(\s+8)!}{(\s+3)!} a(k_1,k_2,k_3) \langle C^1 C^2 C^3 \rangle, \xx
(t^1s^2s^3) &:& 2^9 \frac{(\s+2)(\a_2+2)(\a_3+2)}{(k_1+3)(k_2+1)(k_3+1)}
\frac{\a_1!}{(\a_1-5)!} a(k_1,k_2,k_3) \langle C^1 C^2 C^3 \rangle, \xx
(s^1t^2t^3) &:& 2^9 \frac{(\s+4)\a_2\a_3}{(k_1+1)(k_2+3)(k_3+3)}
\frac{(\a_1+6)!}{(\a_1+1)!} a(k_1,k_2,k_3) \langle C^1 C^2 C^3 \rangle, \\
(s^1 B^2 \bar{B}^3) &:& 2^4 \frac{(\s+2)(\a_1+2)\a_2\a_3}{k_1+1} 
a(k_1,k_2,k_3) \langle C^1 C^2 C^3 \rangle, \xx
(t^1 B^2 \bar{B}^3) &:& 2^4 \frac{(\s+4)\a_1(\a_2+2)(\a_3+2)}{k_1+3} 
a(k_1,k_2,k_3) \langle C^1 C^2 C^3 \rangle, \xx
(\phi^1 s^2 s^3) &:& 2^5 \frac{\s(\s+1)(\a_1-1)(\a_1-2)}{(k_2+1)(k_3+1)}
h(k_1,k_2,k_3) \langle T^1 C^2 C^3 \rangle~. \nonumber
\ee

%%%%%%%%%%%%%%%%%%%%%%%%%%%%%%%%%%%%%%%%%%%%%%%%%%%%%%%%%%%%%%%%%
%                                                               %
%                       REFERENCES                              %
%                                                               %
%%%%%%%%%%%%%%%%%%%%%%%%%%%%%%%%%%%%%%%%%%%%%%%%%%%%%%%%%%%%%%%%%

%\bibliographystyle{h-physrev3}
%\bibliography{bibpp}

\end{document}